\documentclass[epj]{svjour}
\usepackage{latexsym}
\usepackage{graphics}
\begin{document}
\title{A limited resource model of fault-tolerant capability
against cascading failure of complex network }

\author{Ping Li \inst{1}, \ Bing-Hong Wang \inst{2,3}, \ Han Sun \inst{1}, \ Pan Gao \inst{1}, \ Tao Zhou \inst{2,4}}

\offprints{zhutou@ustc.edu (T.Z.)}
\institute{Department of Basic Sciences, Nanjing Institute of
Technology, Nanjing 211167, China \and Department of Modern Physics
and Nonlinear Science Center, University of Science and Technology
of China, Hefei 230026, China\and Shanghai Academy of System
Science, Shanghai, 200093 China \and Department of Physics,
University of Fribourg, CH-1700 Fribourg, Switzerland}
\date{Received: date / Revised version: date}
%
\abstract{We propose a novel capacity model for complex networks
against cascading failure. In this model, vertices with both higher
loads and larger degrees should be paid more extra capacities, i.e.
the allocation of extra capacity on vertex $i$  will be proportional
to $ k_{i}^{\gamma} $, where $ k_{i}$ is the degree of vertex $i$
and $\gamma>0$ is a free parameter. We have applied this model on
Barab\'asi-Albert network as well as two real transportation
networks, and found that under the same amount of available
resource, this model can achieve better network robustness than
previous models.  
} 
\maketitle
\section{Introduction}
\label{intro} Many systems in nature and society can be described by
networks, including biological and social systems, the Internet, the
world-wide web, friendship networks, computer networks, metabolic
networks, power grids, scientific citations, neural networks, and so
on. Therefore, complex networks have recently attracted considerable
attention in physics and other fields. Interestingly, many
real-world networks share a certain number of common topological
properties, such as small-world and scale-free properties
\cite{Newman,Boccaletti,Albert,Dorogovtsev}. The network robustness
is one of the central topics in studying of complex networks.
Robustness refers to the malfunction avoiding ability of a network
when a fraction of its constituents are damaged. Previous works
demonstrated that the heterogeneity of a network induces a high
resilience to random failure while a high sensitivity to intentional
attacks \cite{Jeong,Cohen 1,Cohen 2}. A common failure of many
networks is cascading failure triggered by the removal of vertices
or overload breakdown of vertices. The robustness of complex
networks in response to cascading failure of intentional attacks has
become a topic of recent interest. Prior studies have shown that the
fault-tolerant capability of network have a great impact on their
robustness and function \cite{Jeong,Cohen 1,Cohen 2,Holme 1,Holme
2,Motter-Lai,Motter 2,Crucittia,Zhou3,Lai
1,Liu1,Liu2,Zhao1,Crucitti1,Motter 3,Crucitti 2,Holme
3,Gallos,Latora,Lee,Lai 2,Zhao
2,Kinney,Dall'Asta,Bakke,Tanizawa,Paul,Wang2006,Wang,Sreenivasan,Guo,Centola}.

Any failures of vertices in general will change the distribution of
loads. Here the load of a vertex (also called betweenness centrality
\cite{Freeman,Zhou2006}) is defined as the total number of shortest
paths passing through this vertex. If the load at a particular
vertex increases and becomes larger than its capacity, the
corresponding vertex fails. The overload failure leads to
redistribution about the loads and, as a result, subsequently
cascading failures may occur. Because of the global redistribution
of load, new overload failures may be driven by events happening far
away. This cascading process may stop after a few steps, while it
could also spread over the entire network.

\begin{figure}
\resizebox{0.45\textwidth}{!}{%
  \includegraphics{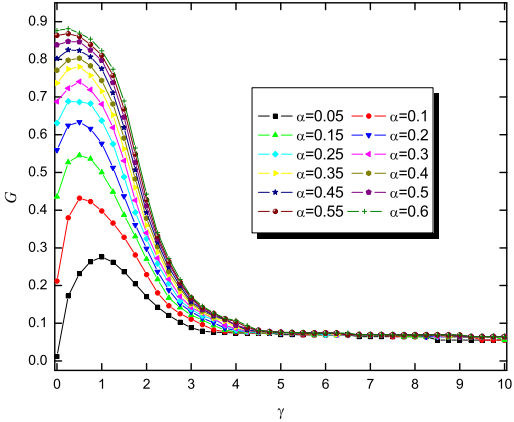}
} \caption{The robustness $G$ vs. $\gamma$ under attacking of the
highest-load vertex on BA network of size $N=5000$ and average
degree $\langle k \rangle =4$ for different $\alpha$.}
\label{fig:1}       
\end{figure}

The damage caused by cascading failures can be quantified by the
relative size of the largest connected component $G$, defined as
following
\begin{equation}
G=N^{\prime}/N,
\end{equation}
where $N$ and $N^{\prime}$ are the numbers of vertices in the
largest component before and after the cascade, respectively. The
integrity of a network is maintained if $G \approx1$, while
breakdown occurs if $G\approx 0$. The relative size $G$ also
represents the robustness of complex networks against cascading
failures. The cascading failure can be prevented by assigning extra
capacities to vertices. Since the extending of network capacity will
bring economic and technique pressure, it is important to explore
how to rationally allocate the limited capacity onto vertices, and
efficiently improve the robustness of network. The capacity of a
vertex is the maximum load that the vertex can handle. Assuming the
capacity $C_{i}$ of vertex $i$ be proportional to its initial load
$L_{i}$:
\begin{equation}
C_{i}=\lambda_{i}\cdot L_{i},
\end{equation}
where $\lambda_{i}>1$ is the tolerance parameter. Excess term
$(\lambda_{i}-1)$ is the extra capacity of vertex $i$, which
simultaneously reflects $i$'s ability of tolerating the additional
burden and the extra cost to protect $i$.

\section{Model}
Among the previous works, two models should be paid special
attention: Motter-Lai (ML) model \cite{Motter-Lai} and Wang-Kim (WK)
model \cite{Wang}. ML model assumes the capacity $C_{i}$ of vertex
$i$ be proportional to the initial load $L_{i}$ as
\begin{equation}
C_{i}=(1+\alpha)\cdot L_{i},\texttt{     }i=1,2,\dots,N,
\end{equation}
where $\alpha\ge0$ is the control parameter representing the extra
capacity. In WK model, the capacity $C_{i}$ is
\begin{equation}
C_{i}=(1+\alpha \cdot \Theta
(\frac{L_{i}}{{L_{\texttt{max}}}}-\beta))\cdot L_{i},i=1,2,\dots,N,
\end{equation}
where $\Theta(x)=0(1)$ for $x<0(>0)$ is a two-valued function,
namely the Heaviside step function,
$L_{\texttt{max}}=\texttt{max}_{i}(L_{i})$, $\alpha\in[0,\infty)$
and $\beta\in[0,1]$ are two control parameters. Here, each vertex
can be in one of two states: assigned some extra capacity, or not.
When $\beta=0$, WK model degenerates to ML model. Since
fault-tolerant capacity is costly
\cite{Jeong,Motter-Lai,Zhao1,Crucitti1,Wang}, a fundamental concern
is how to efficiently allocate limited resources of capacity to
makes network more robust. ML model raises a linear correlation
between extra capacity and initial load, while WK model prefers to
protect the highest-load vertices. We agree with WK model that under
the same cost, a certain more heterogeneous distribution of extra
capacities can further enhance the robustness, however, the two-step
function may be oversimplified. According to the previous works
\cite{Motter-Lai,Wang}, the total cost $e$ to protect a network can
be defined as:
\begin{equation}
e=\frac{1}{{N}}\sum\limits_{i=0}^{N}(\lambda_{i}-1).
\end{equation}
In ML model, the cost is
\begin{equation}
e_{ML}=\frac{1}{{N}}\sum\limits_{i=0}^{N} (\lambda_{i}-1)=
\frac{1}{{N}}\sum\limits_{i=0}^{N}(\alpha)=\alpha .
\end{equation}
In the WK model, the cost is
\begin{equation}
e_{WK}=\frac{1}{{N}}\sum\limits_{i=0}^{N} (\lambda_{i}-1)=
\frac{1}{{N}}\sum\limits_{i=0}^{N}\alpha\cdot
\Theta(\frac{L_{i}}{{L_{\texttt{max}}}}-\beta)=\alpha\cdot
\frac{N^{\prime\prime}}{{N}},
\end{equation}
where $N^{\prime\prime}$ is the number of vertices with initial load
larger than $\beta L_{\texttt{max}}$.

\begin{figure}
\resizebox{0.45\textwidth}{!}{%
  \includegraphics{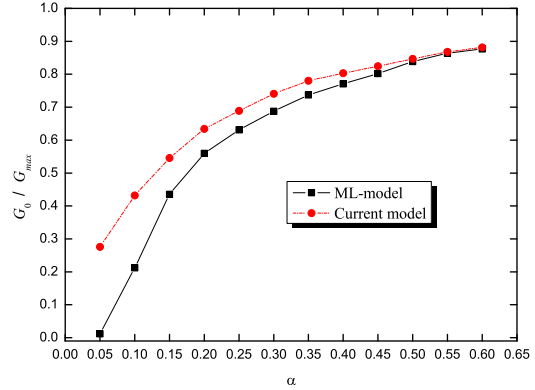}
} \caption{$G_{0}$ and $G_{\texttt{max}}$ vs. $\alpha$.}
\label{fig:2}
\end{figure}

In this paper, we propose a novel allocation mechanism of limited
resources of capacity against cascading failures, as:
\begin{equation}
C_{i}=(1+\alpha k_{i}^{\gamma}/\langle k^\gamma \rangle) \cdot
L_{i},
\end{equation}
where $\alpha\geq 0$ and $\gamma\geq 0$ are two free parameters, and
\begin{equation}
\langle k^\gamma
\rangle=\frac{1}{N}\sum\limits_{i=1}^{N}k_{i}^{\gamma}.
\end{equation}
The parameter $\alpha$ corresponds to the extra capacity, and
$\gamma$ is the parameter controlling the heterogeneity of resource
allocation. the cost of this model is
\begin{equation}
e=\frac{1}{{N}}\sum\limits_{i}^{N}(\lambda_{i}-1)=
\frac{1}{{N}}\sum\limits_{i}^{N}\alpha k_{i}^{\gamma}/\langle
k^\gamma \rangle=\alpha.
\end{equation}

Different from ML model, the fault-tolerant capacity and topological
structure are interrelated in the current model. Since the
components of a network may have different fault-tolerant
capacities, this work raises a more general allocation strategy for
extra capacity. The assigning extra capacity depends not only on
load $L_{i}$ of vertex $i$ but also on the number of its links. When
$\gamma>0$, vertices with larger degrees could be allocated more
extra capacities. The case $\gamma=0$ degenerates to ML model, and
the case $\gamma\to\infty$ represents the extremely heterogeneous
allocation where only the most connected vertex is protected.

\section{Simulation}
Fig. 1 reports the robustness $G$ as a function of $\gamma$ at
several typical values of $\alpha$ for BA network
\cite{Science,Physica} of size $N=5000$ and average degree $\langle
k\rangle =4$. We find that the robustness has a maximum, $G_{max}$,
with $\gamma\in (0,1)$ for different $\alpha$, indicating that the
allocation of the limited resource of fault-tolerant capability
should be neither uniform nor extremely uneven, but a little bit
more heterogenous (for $\gamma>0$) than the case of ML model. When
$\gamma\ge5$, all $G(\gamma)$ at different $\alpha$ converge to a
single curve, indicating that the extremely uneven distribution of
resource, with the largest-degree node possessing the majority of
extra capacity, has even worse performance then the uniform
allocation (except for the cases with $\alpha$ close to zero).
Define $G_0=G(\gamma=0)$, and
$G_{\texttt{max}}=\texttt{max}_{\gamma}G(\gamma)$ for given
$\alpha$. The former corresponds to the ML model, while the latter
to the best robustness predicted by the current model. In Fig. 2, we
plot $G_{0}$ and $G_{\texttt{max}}$ vs. $\alpha$ for the same data
in Fig. 1. Clearly, in a broader region of $\alpha$,
$G_{\texttt{max}}$ is always greater than $G_{0}$, demonstrating
that we can get higher robustness of network than ML model with the
same amount of resource.

\begin{figure}
\resizebox{0.45\textwidth}{!}{%
  \includegraphics{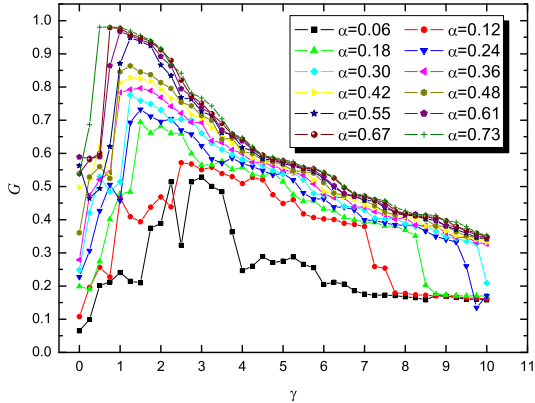}
} \caption{The robustness $G$ vs. $\gamma$ under attacking of the
highest-load vertex for UPTN on Beijing with size $N=4127$ for
different $\alpha$.} \label{fig:3}
\end{figure}

The serious global failure, as a catastrophe resulting from
cascading propagation \cite{Zhou3}, has often been observed in some
kinds of traffic networks suffering overload failure, where the word
``traffic" can stand for information packets of the Internet
\cite{Internet1,Internet2}, the electronic flow in the power grids
\cite{Power}, as well as passengers in public transportation systems
\cite{Zhang2006,Liu2007}. Next, we compare the performance of the
current model with ML model on some real networks. Especially, the
drawing near \emph{Olympic 2008} in Beijing and \emph{World Expo
2010} in Shanghai will bring extremely heavy traffic pressure on
Beijing and Shanghai urban traffic systems, thus an urgent problem
is how to allocate the limited additional traffic capacities to
avoid possible cascading traffic congestions. The topological
characters of urban public traffic networks (UPTNs) are reported
previously \cite{Li2006}, and here we investigate the same data to
see if our model can achieve better performance than ML model.

\begin{figure}
\resizebox{0.45\textwidth}{!}{%
  \includegraphics{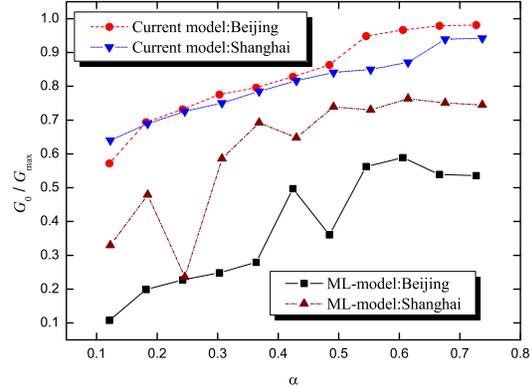}
} \caption{$G_{0}$ and $G_{\texttt{max}}$ vs. $\alpha$. for UPTNs in
Beijing ($N=4127$) and Shanghai (N=2035).} \label{fig:5}
\end{figure}

In Fig.3, we show the relation between the robustness $G$ of
Beijing's UPTN and the parameter $\gamma$ of allocation of capacity
resources at various values of $\alpha$. Qualitatively speaking, the
pattern of $G(\gamma)$ is similar to that of BA network. Fig. 4
compared the performance of the current model and ML model for UPTNs
in Beijing and Shanghai respectively. Clearly, the current model can
always perform better than ML model under the same amount of extra
capacity. Moreover, large fluctuation is observed for ML model, thus
one could not guarantee the enhancement of robustness giving more
resource. In contrast, $G_{\texttt{max}}$ increases with $\alpha$
monotonously in the current model. Two $G(\alpha)$ curves in Beijing
and Shanghai are remarkably different for ML model, while almost
have almost the same trend for the current model. Perhaps, this
indicates that current model has some universality to improve
robustness for different real-world networks, or maybe it is because
the robustness of this strategy is near the optimal one.

\section{Summary}
In summary, the robustness of complex networks is analyzed within
the framework of allocating extra fault-tolerant capacity against
cascading failure. We proposed a novel capacity model, which
achieves higher robustness for the same amount of resource as ML
model \cite{Motter-Lai}. In previous works, the extra capacity of a
vertex is linearly correlated with its load, or simply divided into
two discrete values by a step function. In this paper, we argue that
those prior strategies may be oversimplified, and a more complicated
and more heterogenous (than ML model) allocation strategy could
further enhance the robustness. The simulation results of BA
networks and two typical real transportation networks strongly
support the efficiency of our strategy. It is worthwhile to
emphasize two features of the current model: Firstly, the maximum
value of robustness $G_{\texttt{max}}$ is a monotonously increasing
function on parameter $\alpha$, in contrast with ML model. Secondly,
the same trend and pattern of $G_{\texttt{max}}(\alpha)$ for
different UPTNs indicates that this model may have some universality
for robustness improvement of different real-world networks (Maybe
it is because the performance of this strategy is near the optimal
one). However, this point is not clear thus far, which needs further
investigation. Besides, although the previous works suggested using
the mean relative extra capacity $e$ to measure the cost, it is
also, to some extent, reasonable to measure the cost by using
absolute extra capacity. In that way, given $\alpha$, the cost will
be a monotonous function of $\gamma$, and our numerical results give
raise to a practically significant conclusion: There exists a
specific threshold of $\gamma_c$ corresponding to
$G_{\texttt{max}}$, below which one could improve the robustness by
spending more, while above which the additional resource will, in
contrast to what we expect, makes the network worse robust. Although
at a first step far from the final goal, optimal allocation, we
believe this model have its theoretical importance and potential
application in designing infrastructure networks from the point of
economic view. It can also provide guidance in designing more robust
artificial networks.

This work is funded by the National Basic Research Program of China
(973 Project No. 2006CB705500), the National Natural Science
Foundation of China (Grant Nos. 10472116, 10532060, and 10635040),
and the Science and Technology Foundation (KXJ06048) as well as the
Student Foundation (NB2006183) of Nanjing Institute of Technology.

\end{document}